\documentclass[twocolumn,apl,amsmath,amssymb,showpacs,superscriptaddress]{revtex4-2}
\usepackage{epsf}      
\usepackage{graphicx}
\usepackage{color}
\usepackage{textcomp, gensymb}
\usepackage{amsmath}
\usepackage{array}
\usepackage{float}
\usepackage[hidelinks,colorlinks=true,linkcolor=blue,citecolor=blue]{hyperref}

\begin{document}

\title{Structural, vibrational and electronic properties of Nb substituted orthovanadates LaV$_{1-x}$Nb$_x$O$_4$}

\author{Ashok Kumar}
\affiliation{Department of Applied Physics, Delhi Technological University, Delhi-110042, India}
\affiliation{Department of Physics, Atma Ram Sanatan Dharma College, University of Delhi, New Delhi-110021, India}
\author{Anurag Sharma}
\affiliation{Department of Physics, Indian Institute of Technology Delhi, Hauz Khas, New Delhi-110016, India}
\author{Madhav Sharma}
\affiliation{Department of Physics, Indian Institute of Technology Delhi, Hauz Khas, New Delhi-110016, India}
\author{Vinod Singh}
\affiliation{Department of Applied Physics, Delhi Technological University, Delhi-110042, India}
\author{Anita Dhaka}
\affiliation{Department of Physics, Hindu College, University of Delhi, New Delhi-110007, India}
\author{Rajendra S. Dhaka}
\email{rsdhaka@physics.iitd.ac.in}
\affiliation{Department of Physics, Indian Institute of Technology Delhi, Hauz Khas, New Delhi-110016, India}

\date{\today}                          

\begin{abstract}

We investigate the structural, vibrational, morphological, and electronic properties of Nb substituted orthovanadate LaV$_{1-x}$Nb$_x$O$_4$ samples prepared by the solid-state reaction method. The x-ray diffraction (XRD) analysis reveals the presence of three crystal structures [monoclinic monazite ($m-m$) type for the $x=$ 0, two-phase equilibrium of monoclinic monazite ($m-m$) and tetragonal scheelite ($t-s$) type for the 0.2$\leq$$x$$\leq$0.8, and monoclinic fergusonite ($m-f$) type for the $x=$ 1 samples] with an increase in Nb$^{5+}$ concentration. The Raman spectroscopy and x-ray photoelectron spectroscopy (XPS) were employed to study the vibrational and electronic properties of all the samples, respectively. In order to choose an excitation wavelength that does not cause undesirable fluorescence and has observable intensities of all the vibrational modes, the Raman spectra are collected using 532 nm, 633 nm, and 785 nm laser lines. With increasing the Nb$^{5+}$ concentration, new Raman modes associated with Nb-bonds are clearly visible and the intensity of V-bonds assigned modes is decreasing. The XPS analysis shows the unchanged 3+ oxidation state of La ion where the intensity of the V 2$p$ core-level decreases while the Nb 3$d$ core-level increases with $x$. The equal spin-orbit energy splitting of the states is confirmed by the average energy difference (across La core-level spectra for all the samples) for state I as well as bonding and anti-bonding of state II. Interesting, the relative intensity of La 3$d$ state I and state II show systematic change with Nb doping altering the metal ligand overlap. We discuss and provide insight into the evolution of the structural, morphological, and chemical features with Nb substitution in LaV$_{1-x}$Nb$_x$O$_4$ samples. 
\end{abstract}

\maketitle

\section{\noindent ~Introduction}

In various polycrystalline oxides, rare earth orthovanadates (RVO$_4$; R-Rare earth elements) are interesting because of their potential applications in catalysis, polarizers, luminescent materials, and laser host materials \cite{Varghese_PRB_20, Huang_JALCOM_19, Carbonati_JALCOM_21}. Also, researchers have reported that complex oxide materials show interesting structural, magnetic and electronic properties \cite{Kumar_JPCL_22, Kumar_PRB3_22, Kumar_PRB1_20, Kumar_PRB2_20}, and may be utilized for various applications such solid oxide fuel cells and as an electrode material for Lithium-ion batteries due of their high specific capacity and cycle stability \cite{YiJAL17}. It is interesting to note that the lanthanum based orthovanadates LaVO$_4$ shows the structural trend in rare-earth family, it crystallizes in tetragonal--zircon ($t-z$) type polymorphs with space group I4$_1$/amd and monoclinic--monazite ($m-m$) type polymorph with space group P2$_1$/n. However, it thermally stabilizes in $m-m$ type, whereas the $t-z$ structure remains in metastable state at room temperature, because of the largest ionic radius of La$^{3+}$ in lanthanide series, it has a higher oxygen coordination number (9) in $m-m$ type structure as compared to 8 in $t-z$ type \cite{SunJAP10}. The zircon structure contains a pattern of VO$_4$ tetrahedra (having four identical V-O bonds) \cite{ChakoumkosJSSC94} and RO$_8$ dodecahedra (coordination no. 8), sharing their edges alternatively and linked together in chains along the $c-$axis. In the monazite structure, deformed VO$_4$ tetrahedra with four different V-O bonds \cite{RiceACB76} are connected to RO$_9$ polyhedra (coordination no. 9) and sharing their edges. The zircon type LaVO$_4$ sample is difficult to prepare at ambient conditions by conventional solid state reaction method but few reports say that it can be synthesized and stabilized by hydrothermal and precipitation methods \cite{FanJPCB06, RastogiJPCC17, Xie_JALCOM_12}. 

The structural and electronic properties of lanthanum orthovanadate with pentavalent niobium substitution are vital to understand for their practical use. Though the parent compound  LaVO$_4$ with substitution at the La site has been extensively explored \cite{suzukiJAC14, LiuJSSC12}, there are very few studies to understand the effect of substitution at V site \cite{VermaACAG01, HimanshuPRB21}. As the niobium is located just below vanadium in the periodic table and has many advantages like Vanadium prices have recently risen to about 300\% higher, niobium (Nb$^{5+}$) is biocompatible, isoelectronic to vanadium ion and has a larger ionic radius (0.48~\AA) with four coordination numbers in comparison to vanadium ion (0.36~\AA) \cite{ErrandoneaPMS08}. The LaNbO$_4$ is a rare-earth niobate and shows a well-known temperature and composition/substitution-induced structural transformation. For example, the LaNbO$_4$ undergoes a thermally induced structural transition from monoclinic fergusonite ($m-f$) with space group I2/a to tetragonal scheelite ($t-s$) with space group I4$_1$/a) phase at $\sim$495$\degree$C \cite{TakeiJCG77}. Similarly, it undergoes structural transformation by substituting Nb$^{5+}$ at V$^{5+}$ site \cite{AldredML83}. It has been reported that lanthanum niobate shows interesting properties and very useful for technological applications such as proton conductivity \cite{HaugsrudNM06, Hakimova_CI_19}, good dielectric, high energy emission using X-ray excitation \cite{BlasseCPL90} and its potential for applications in a variety of fields, including sensors \cite{Liu_SABC_22}, contrast agents, waveguides, ferroelectrics \cite{Zhou_ICF_21}, phosphors \cite{Xue_CEJ_21}, laser crystals \cite{DingRSCA17}, luminophores, LEDs \cite{Xiong_APA_20}, etc.

In this paper, we study the structural, vibrational, morphological, and electronic properties of LaV$_{1-x}$Nb$_x$O$_4$ using various experimental tools like x-ray powder diffraction (XRD), scanning electron microscopy (SEM), high resolution transmission electron microscopy (HR-TEM), selected area electron diffraction (SAED), Raman spectroscopy, and x-ray photoelectron spectroscopy (XPS). We find the phase purity and structural transition by performing the Rietveld refinement of the measured XRD patterns at room temperature. The Raman spectra of LaV$_{1-x}$Nb$_x$O$_4$ samples are measured with different excitation wavelengths of 532 nm, 633 nm, and 785 nm, where we find significant intensity of all the Raman active modes as well as interesting changes with Nb substitution. The Raman spectra exhibit a pattern of maximum intensity peaks that is compatible with Badger's rule. The structural phase transition observed in the XRD analysis of LaV$_{1-x}$Nb$_x$O$_4$ is also supported by the intensity variation in the Raman mode observed in the samples with increasing Nb concentration. Through the SEM micrographs, we identify that the samples contain fine particles along with pores as well as  changes in particle size and shape can be seen in the surface images of the samples. The core-level photoemission reveals the oxidation state and electronic structure of the constitute elements in these samples. The intensity of the core-level spectra of all the samples varied systematically with an increase in Nb$^{5+}$ concentration, as shown by XPS analysis. The average energy difference (for the La core-level spectra of all the samples) for state I, state II bonding, and state II anti-bonding verified the equal spin-orbit energy splitting of the states. Moreover, we find a systematic change in the relative intensity of La 3$d$ state I and state II with Nb doping, which suggest an altering in the metal ligand overlap.

\section{\noindent ~Experimental}

We use solid-state reaction method to prepare LaV$_{1-x}$Nb$_x$O$_4$ ($x=$ 0 to 1)  samples by mixing V$_2$O$_5$ (99.6$\%$, Sigma), Nb$_2$O$_5$ (99.99$\%$, Sigma), and La$_2$O$_3$ (99.99$\%$, Sigma) as precursors in the stoichiometric proportions. The La$_2$O$_3$ was pre-dried for 6 hrs at 900$\degree$C to remove the moisture. After that the mixture was ground evenly for 8 hours, then heated for 17 hrs at 1000$\degree$C. The mixture was then reground and sintered at 1250$\degree$C for 13 hrs to improve the crystallinity of the samples. The phase purity and structural parameters of LaV$_{1-x}$Nb$_x$O$_4$ were determined using Panalytical XPert$^3$ powder x-ray diffractometer at room temperature using the Cu source of K$\alpha$ radiation ($\lambda$ = 1.5406~\AA~). We use the step size of 0.033$\degree$ for each XRD scan taken in the 2$\theta$ range from 10$\degree$ to 90$\degree$. The lattice parameters are extracted by the Rietveld refinement of XRD patterns using FullProf software, where linear interpolation is used to fit the background. We use Jeol JSM-7800F Prime field emission scanning electron microscope (FE-SEM) with LN$_2$ free SDD X-max 80 EDS detector in high vacuum mode to produce the scanning electron microscope (SEM) micrographs of the materials' surfaces. The analysis of particle size and change in morphology of LaV$_{1-x}$Nb$_x$O$_4$ was done using ImageJ software by analyzing SEM micrographs at the surface of the pellet samples. In order to execute FE-SEM, the non-conducting LaV$_{1-x}$Nb$_x$O$_4$ pellets were turned into conducting by coating the surface with a thin layer of Au using a sputter coater. We use the JEOL/JEM-F200 microscope, equipped with thermal electron field emission and OneView CMOS camera (4k $\times$ 4k pixels), to collect HR-TEM data by operating the system at an acceleration voltage of 200~keV. 

The Raman spectra were recorded at room temperature with the Renishaw inVia confocal Raman microscope using 2400~lines/mm grating, 10X objective, and three different wavelengths; (i) 532~nm, gas laser with a power of 1~mW, (ii) 633~nm, where the semiconductor diode laser with a power of 1~mW, (iii) 785~nm semiconductor diode laser with a power of 0.1~mW. The samples can be identified by their particular Raman fingerprint, and their structural and chemical information can be discovered through the examination of several Raman active modes in LaV$_{1-x}$Nb$_x$O$_4$. The x-ray photo emission spectroscopy (XPS) measurements are done using AXIS Supra instrument (Kratos Analytical Ltd). The survey spectra and core level spectra (La 3$d$, Nb 3$d$, V 2$p$, and O 1$s$, for each sample), were recorded at room temperature using the monochromatic X-ray source: Al K$\alpha$-1486.6 eV(step size 1~eV for the survey and 0.1~eV for core level spectra), with a charge neutralizer, is used to offset the charging impact in these insulating materials. The pass energy of the analyzer was 160~eV and 20~eV for the survey and core-level spectra, respectively. For all the wide scans and core-level spectra, the C 1$s$ peak is fitted to obtain the peak binding energy (BE) and the calibration for charge correction was done using the C 1$s$ BE reference at 284.6~eV for each sample. We utilize the Igor Pro 9 software to analyze the observed Raman spectra and fitted the modes using the Lorentzian peak function as well as XPS spectra using the Voigt function.

\section{\noindent ~Results and Discussion}

\begin{figure*}
\includegraphics[width=6.8in, height=7 in]{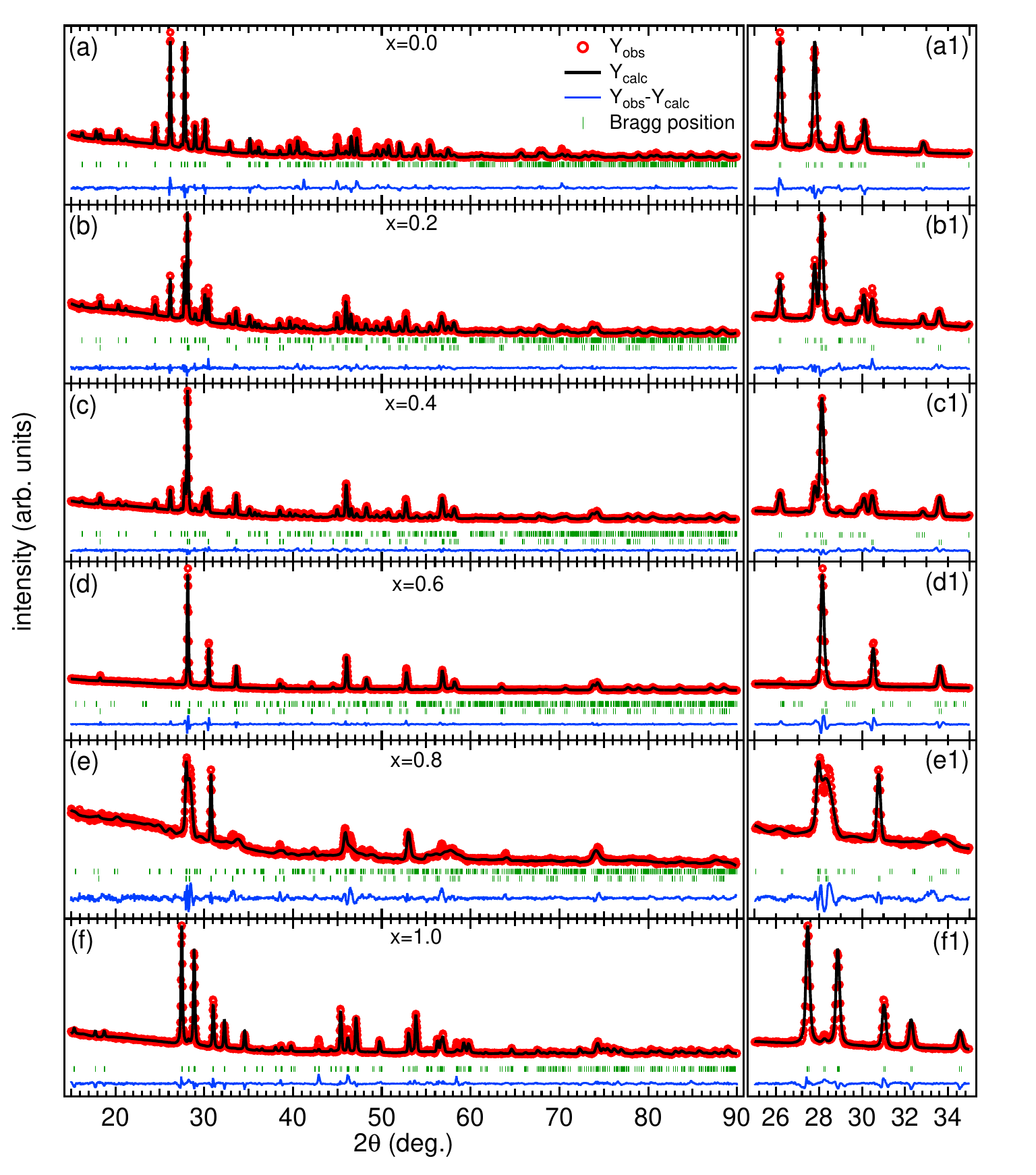}
\caption{(a--f) The Rietveld refined x-ray diffraction patterns of LaV$_{1-x}$Nb$_x$O$_4$ ($x=$ 0--1) samples. The experimental, simulated, and difference between the experimental and simulated spectra are shown by open red circles, black solid lines, and blue solid lines, respectively. The Bragg positions corresponding to their respective space groups are shown by green vertical markers. In the side of each panel (a1--f1), we show partial amplification for clarity between 2$\theta$ = 25--35$\degree$ for all the samples.}
\label{fig:XRD}
\end{figure*}

The Rietveld refined room-temperature x-ray diffraction (XRD) patterns of the polycrystalline LaV$_{1-x}$Nb$_x$O$_4$ ($x=$ 0--1) samples are displayed in Fig.~\ref{fig:XRD} and lattice parameters of the samples are summarised in Table \ref{tab:XRD}, where we can see that angle $\beta$ is increasing in the $m-m$ type phase of LaV$_{1-x}$Nb$_x$O$_4$ samples with Nb$^{5+}$ substitution due to higher ionic size of Nb$^{5+}$ as compared to V$^{5+}$. The crystallization of LaV$_{1-x}$Nb$_x$O$_4$ is clearly observed in three different phases depending on the substitution of Nb$^{5+}$ at the site of V$^{5+}$, and also been reported by Aldred \textit{et al.} in ref.~\cite{AldredML83}. We observe that the structure changes from $m-m$ to $m-f$ with increase in the Nb$^{5+}$ concentration from 0 to 100\%. For the $x=$ 0 and 1, a pure monoclinic phase is obtained with no impurity peaks. In between $x=$ 0.2 and 0.8, a monoclinic monazite ($m-m$) and a tetragonal scheelite ($t-s$) type phases coexist. Moreover, all the Bragg reflections of LaVO$_4$ and LaNbO$_4$ can easily be indexed to the $m-m$ and $m-f$ phases with the space group P2$_1$/n and I2/a for the $x=$ 0 and 1 samples, respectively. We find that the contribution of space group I4$_1$/a is increasing from the $x=$ 0.2 to 0.8 samples (see Table \ref{tab:XRD}) due to the increase of $t-s$ phase with the substitution of Nb$^{5+}$ at the site of V$^{5+}$ in LaV$_{1-x}$Nb$_x$O$_4$ samples. So, it can clearly be seen that the LaV$_{1-x}$Nb$_x$O$_4$ samples crystallize in monoclinic monazite ($m-m$) type ($x=$ 0), coexistence of monoclinic monazite ($m-m$) and tetragonal scheelite ($t-s$) type (0.2$\leq$$x$$\leq$0.8), and monoclinic fergusonite ($m-f$) type ($x=$ 1) \cite{AldredML83, SunCI15}.

\begin{table*}
		\caption{The Rietveld refinement parameters of polycrystalline LaV$_{1-x}$Nb$_x$O$_4$ ($x=$ 0--1) samples with Nb substitution induced metastable tetragonal--scheelite phase for the $x=$ 0.2 to 0.8 samples, determined using the FullProf software.}
		\centering 
		\begin{tabular}{|c|c|c|c|c|c|c|c|}
			\hline
			\rule{0pt}{12pt}$~~~x~~~$ &~~~$\chi^2$~~~&~Space ~Group~&\textit{a} (\AA)& \textit{b} (\AA) & \textit{c} (\AA) &$\beta (\degree)$& ~Volume~$(\rm \AA^3)$~\\[0.5ex]
			\hline
			
			\rule{0pt}{12pt}0 & 1.09&P2$ _1 $/n&~~7.042(3)~~&~~7.276(4)~~&6.724(7) &~104.88 (6)~ & 333.033(5)\\ 
			\hline
			\rule{0pt}{12pt}0.2&2.63 &P2$_1$/n - 84$\%$& 7.046(1) & 7.278(2)&6.733(3) &104.91(1) & 333.685(4)\\
			&&I4$_1$/a - 16$\%$&5.336(1)&5.336(1)&11.731(2)&90&334.042(4)\\
			\hline
			\rule{0pt}{12pt}0.4 &2.46 &P2$ _1 $/n - 78$\%$& 7.043(0) & 7.276(3) &6.732(4) & 104.91(2) & 333.397(3)\\
			&&I4$ _1 $/a - 22$\%$&5.332(3)&5.332(3)&~~11.735(2)~~&90&333.509(6)\\
			\hline	
			\rule{0pt}{12pt}0.6& 3.70&P2$_1$/n - 45$\%$&6.818(9)  &7.596(5)&8.030(0)& 105.21(7)  &401.383(6) \\
			&&I4$ _1 $/a - 55$\%$& 5.329(8)& 5.329(8)&11.714(8)&90&332.787(0)\\
			\hline
			\rule{0pt}{12pt}0.8 &4.31 &~~P2$ _1 $/n - 4$\%$~~& 6.878(5) & 7.459(4) &7.679(8) & 105.61(1) & 379.517(2)\\
			&&I4$ _1 $/a - 96$\%$&5.375(4)&5.375(42)&11.624(0)&90&335.869(7)\\
			\hline
			 \rule{0pt}{12pt}1 & 4.91&I2/a&~~5.558(5)~~&~~11.529(1)~~&5.201(8) &~93.99(2)~ & 332.546(3)\\ 
			\hline
	
		\end{tabular}
		\label{tab:XRD}
		
	\end{table*}

Moreover, for the $x=$ 0 sample, the $m-m$ type crystal structure shows high intensity diffraction peaks corresponding to (200) as well as (120) crystal planes at 26.17$\degree$ and 27.78$\degree$, respectively. However, the $t-s$ type structure contains a peak corresponding to (112) plane at 28.08$\degree$, and the $m-f$ type structure shows high intensity peaks for the ($\overline{1}$21) and (121) planes at 27.5$^o$ and 28.9$^o$ respectively. In the measured XRD pattern for the $x=$ 0.2 to 0.8 samples, the diffraction peaks for the (200), (120) and (112) planes are present, which clearly indicate the co--existence of both the $m-m$ or $t-s$ type structures. The presence of (110) plane at 17.65$\degree$ for the $x=$ 0.2 and 0.4 samples is due to the dominance of $m-m$ type structure in the LaV$_{1-x}$Nb$_x$O$_4$. The (200) and (120) peaks are also present in these samples; however, their intensity decreasing with higher concentration of Nb substitution and become negligible for the $x\ge$ 0.6 sample. As the Nb$^{5+}$ concentration becomes more than V$^{5+}$ concentration the $t-s$ type structure dominates, which results in the reduction/absence of diffraction peaks corresponding to (200) and (120) planes. The variation in peak intensity corresponding to (200) and (120) crystal planes and the presence of (112) plane indicate the co--existence of $t-s$ and $m-m$ type structures for the $x=$ 0.2 to 0.8 samples. This also validates that the $m-m$ type structure (P2$_1$/n) is decreasing and $t-s$ type structure (I4$_1$/a) is increasing with increasing the Nb concentration, i.e., from the $x=$ 0.2 to 0.8 samples. The determined $\%$ of phases by Rietveld refinement of XRD data is presented in Table \ref{tab:XRD}. For the $x=$ 1 sample, the presence of ($\overline{1}$21) and (121) peaks further confirms the $m-f$ type structure of LaNbO$_4$ and consistent with literature \cite{Wachowski}. 

Note that pure $m-m$ and $m-f$ phases are observed for the $x=$ 0 and 1 samples, respectively. However, for the $x=$ 0.2--0.8 samples, both the monoclinic and scheelite-tetragonal phases coexist in a certain ratio. These results reveal that LaV$_{1-x}$Nb$_x$O$_4$ samples undergo three phase transformation; monoclinic monazite ($m-m$) type (for the $x=$ 0), two-phase equilibrium of monoclinic monazite ($m-m$) and tetragonal scheelite ($t-s$) type (0.2$\leq$$x$$\leq$0.8), and monoclinic fergusonite ($m-f$) type (for the $x=$ 1) with increased substitution of Nb$^{5+}$ at the V$^{5+}$ site. It is quite interesting to note that small amount of Nb$^{5+}$ substitution can transform LaVO$_4$ from $m-m$ phase to mix of $m-m$ and $t-s$ phases. It has also been observed that LaNbO$_4$ shows structural transition from monoclinic to a tetragonal phase at $\sim$495$\degree$C. This structural transformation is very important in governing the protonic conductivity of LaNbO$_4$ \cite{HuseJSSC12}. For some compositions of LaV$_{1-x}$Nb$_x$O$_4$, this transition temperature shifts near room temperature. The reported temperature-dependent XRD measurements also suggest that at $x=$ 0.75 (25$\%$ substitution of V$^{5+}$ at Nb$^{+5}$ sites in LaNbO$_4$) \cite{AldredML83}, it possess a tetragonal structure at room temperature as its transition temperature is 250~K. The XRD pattern below 250~K shows some residual intensity (broadened lines) of tetragonal structures because of precursor effects. Similarly, we can see broad peaks in XRD patterns for the $x=$ 0.8 sample due to the above mentioned effect \cite{AldredML83}. As we increase the Nb concentration, we find some new peaks appearing in the $x=$ 0.2 sample at 33.56$^o$, 52.68$^o$, 56.69$^o$, and 58.06$^o$. All these peaks are the symbols of t--s structure that belongs to (020), (116), (312), and (224) planes, respectively \cite{DavidMRB83}. These peaks maintained up to the $x=$ 0.8 sample, which confirms presence of some $t-s$ phase, and also indicate the substitution-induced phase transformation. This is an important finding that LaNbO$_4$ can possess a tetragonal structure at room temperature by just 20$\%$ replacement of Nb$^{5+}$ sites by the V$^{5+}$ sites. This result opens the possibility for a wide range of applications of LaNbO$_4$ at room temperature. All these patterns discussed above suggest that substitution of larger Nb$^{5+}$ ($r=$ 0.48 \AA) ion for V$^{5+}$ ($r=$ 0.36 \AA) affects the lattice constant of LaV$_{1-x}$Nb$_x$O$_4$ and confirms the transformation of 3 different phases with increasing concentrations of Nb$^{5+}$. 

\begin{figure}[ht]
\includegraphics[width=3.45in]{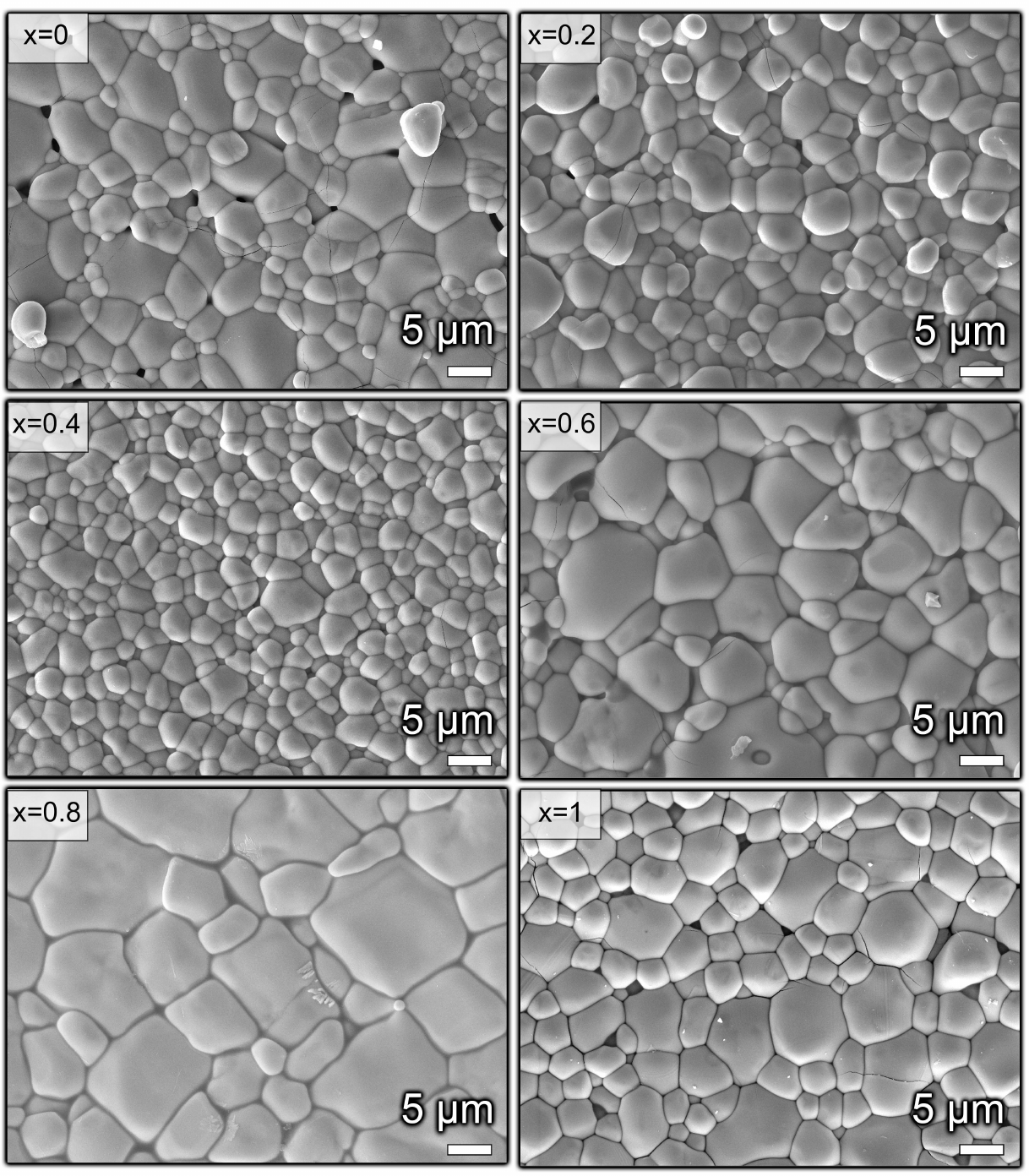}
\caption{The scanning electron microscope images of the LaV$_{1-x}$Nb$_x$O$_4$ ($x=$ 0--1) samples.}
\label{fig:SEM} 
\end{figure}

The scanning electron microscope images of the LaV$_{1-x}$Nb$_x$O$_4$ for the $x=$ 0--1 are shown in Fig.~\ref{fig:SEM}, which depict the closed packed surface morphology in all the samples and some variation in the particle size is clearly visible. The pores are clearly visible from the top view of the surface. We can see that with the increase in Nb$^{5+}$ concentration, the particle size slightly decreases from $x=$ 0 to $x=$ 0.4 sample, then increases and becomes maximum at $x=$ 0.8 and again decreases for the $x=$ 1 sample. 
\begin{figure}[ht]
\includegraphics[width=3.42in]{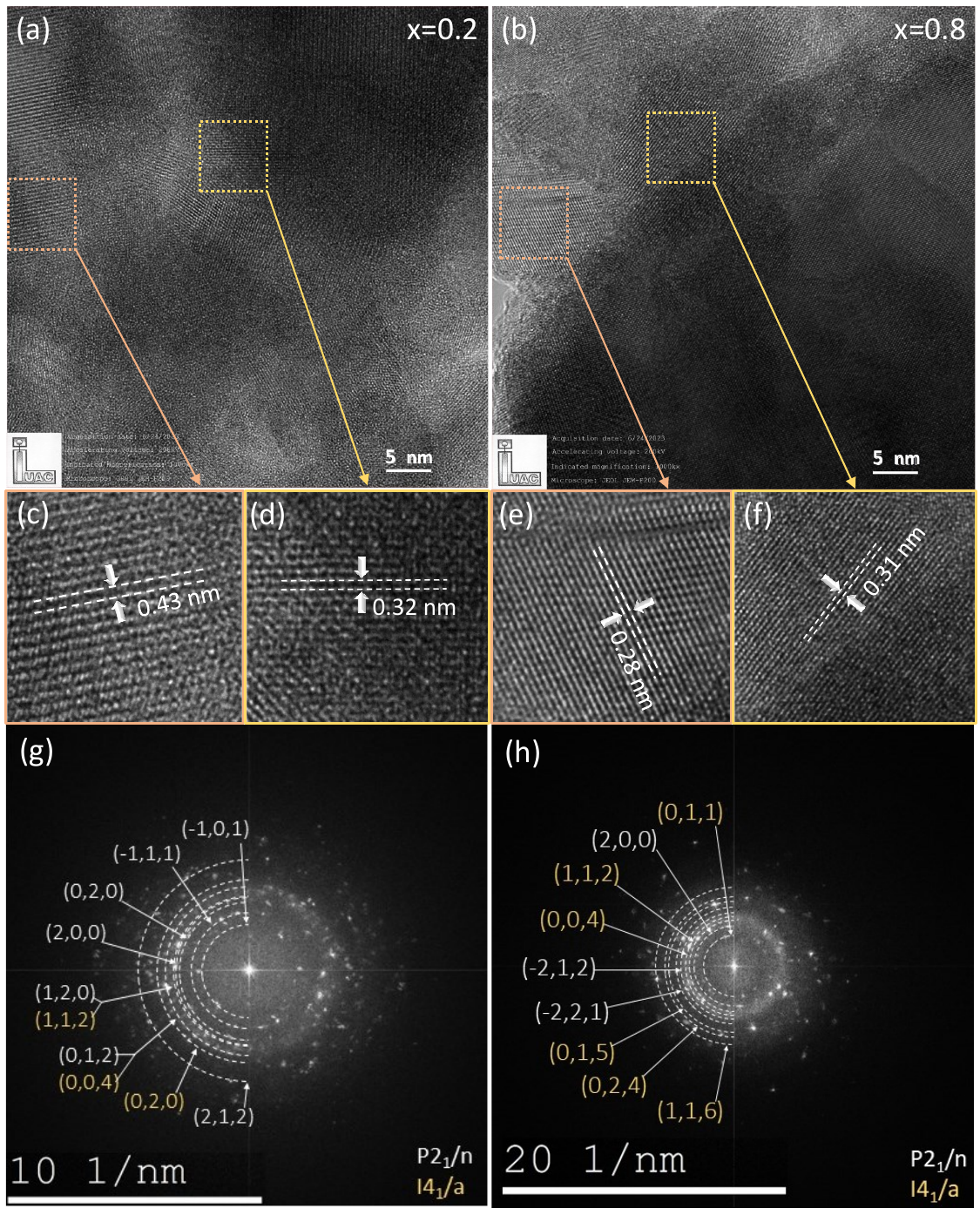}
\caption{The HR-TEM images of LaV$_{1-x}$Nb$_x$O$_4$ for the (a) $x=$ 0.2 and (b) $x=$ 0.8 samples. The magnified view of HR-TEM images in (c, d) for the $x=$ 0.2 sample, and (e, f) for the $x=$ 0.8 sample. (g, h) The SAED patterns for the $x=$ 0.2 and 0.8 samples, respectively.}
\label{TEM} 
\end{figure}
An average particle size (D) of LaV$_{1-x}$Nb$_x$O$_4$ is 5.14 $\micro$m for the $x=$ 0, 4.22 $\micro$m for the $x=$ 0.2, 3.56 $\micro$m for the $x=$ 0.4, 8.73 $\micro$m for the $x=$ 0.6, 11.31 $\micro$m for the $x=$ 0.8, and 5.70 $\micro$m for the $x=$ 1 samples. It is found that the change in crystal surface morphology of LaV$_{1-x}$Nb$_x$O$_4$ samples with increasing Nb$^{5+}$ concentration causes variation in the particle size and shape.

Further, in Figs.~\ref{TEM} (a, b) we display the HR-TEM images indicating distinct sets of planes with characteristic spacing for the $x=$ 0.2 and 0.8 samples. The images in Figs.~\ref{TEM}(c, d) and (e, f) for the samples $x=$ 0.2 and $x=$ 0.8, respectively, show these plane sets in magnified view. The spacing between the planes is determined using ImageJ software, and we find the $d-$spacings of 0.43 and 0.32 nm for the (-1,1,1) and (1,2,0) planes in the $P2_1/n$ phase for the $x=$ 0.2 sample, and 0.28 and 0.31 nm for the (0,0,4) and (1,1,2) planes in the $I4_1/a$ phase for the $x=$ 0.8 sample. However, these planes only correspond to the dominating phase of the mixed-phase samples. The selected area electron diffraction (SAED) patterns in Figs.~\ref{TEM}(g, h) indicate the contributions from both the phases. The indexed ($h, k, l$) planes that relate to $P2_1/n$ are coloured in white and yellow colour is designated to the $I4_1/a$ space group, as marked in Figs.~\ref{TEM}(g, h). We find that the analysis of HR-TEM and SAED results is consistent with the XRD refinement data for these samples, as presented in Figs.~1(b, e). 

\begin{figure*}
\includegraphics[width=2.3in]{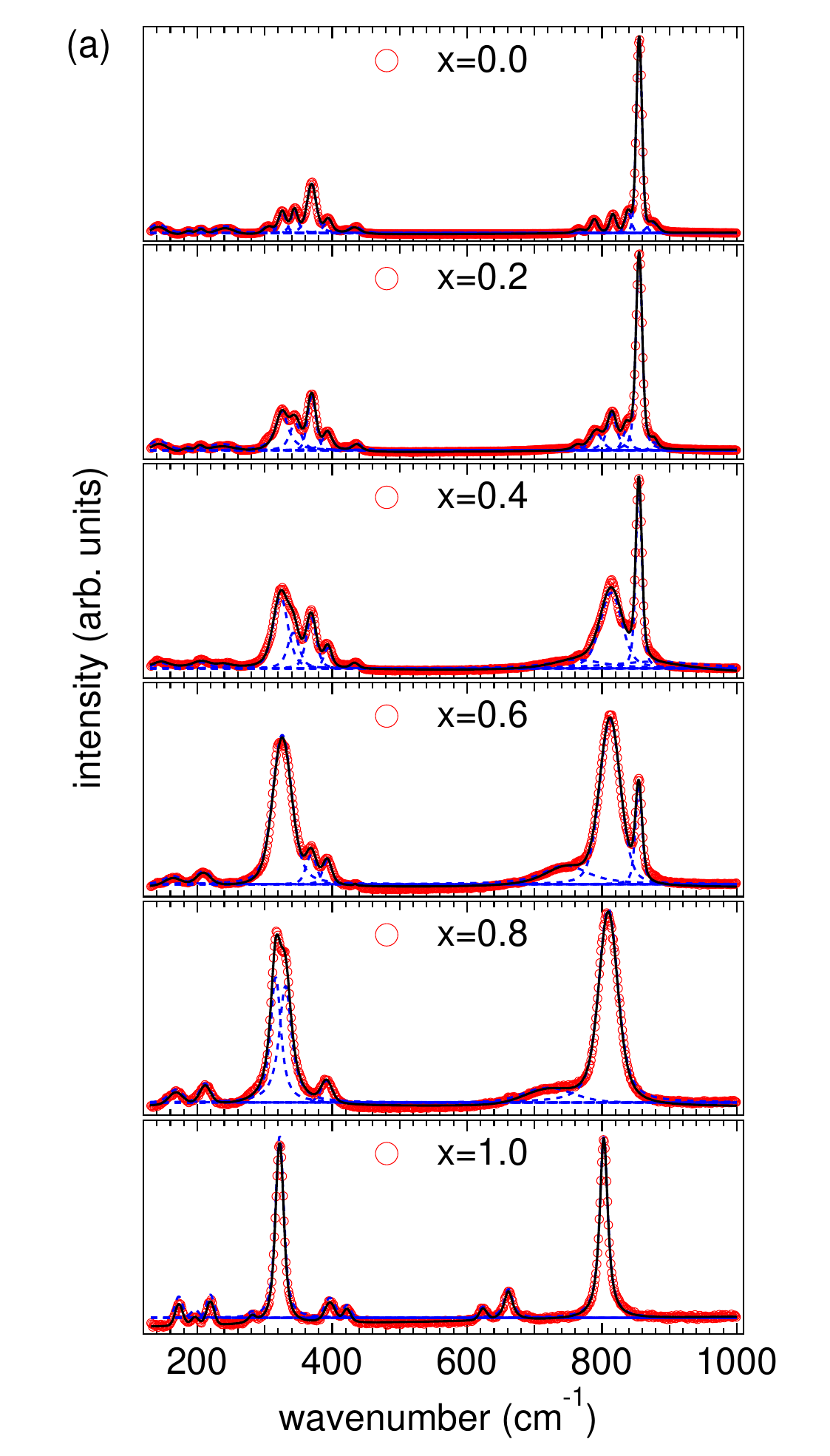} 
\includegraphics[width=2.35in]{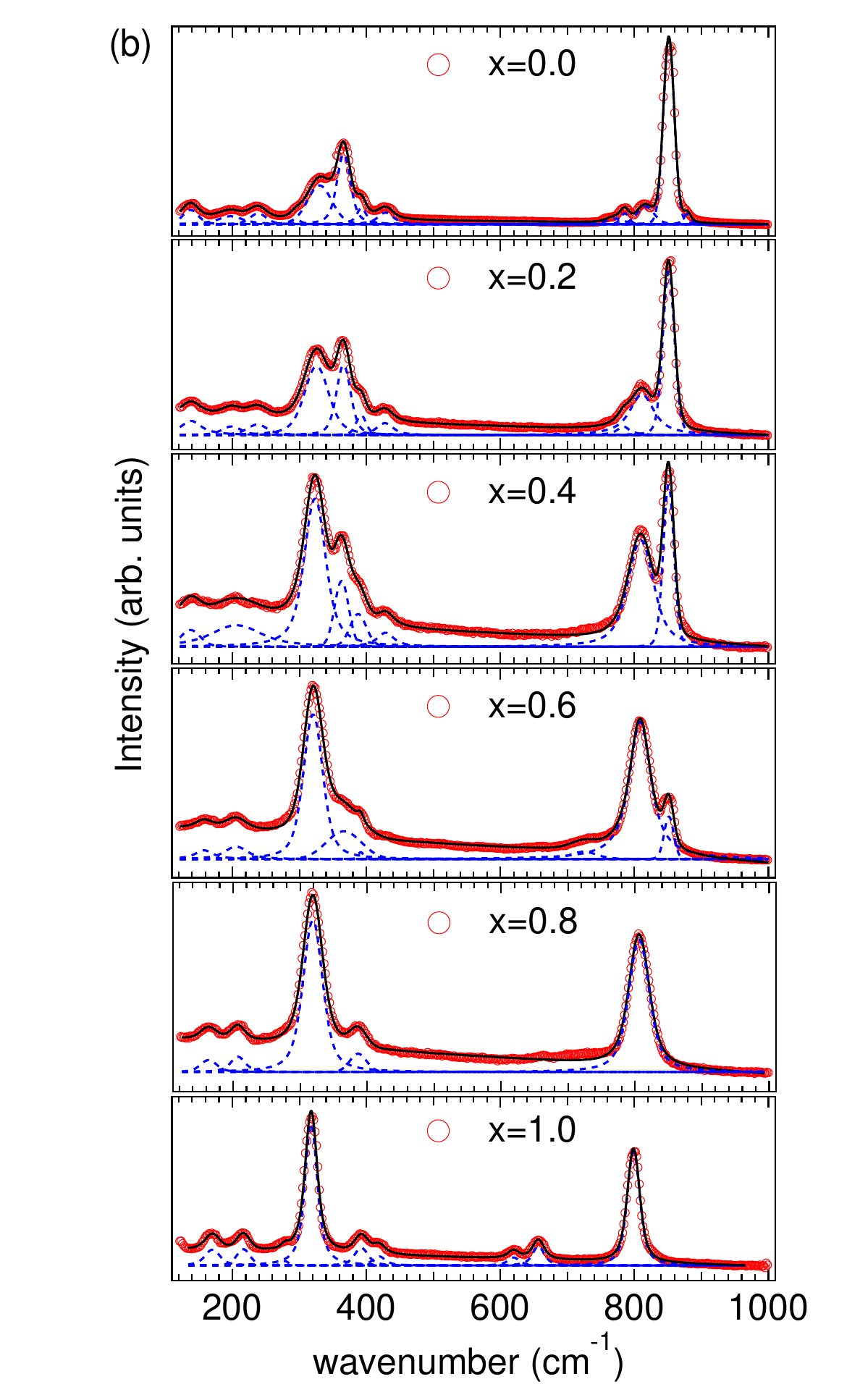} 
\includegraphics[width=2.3in]{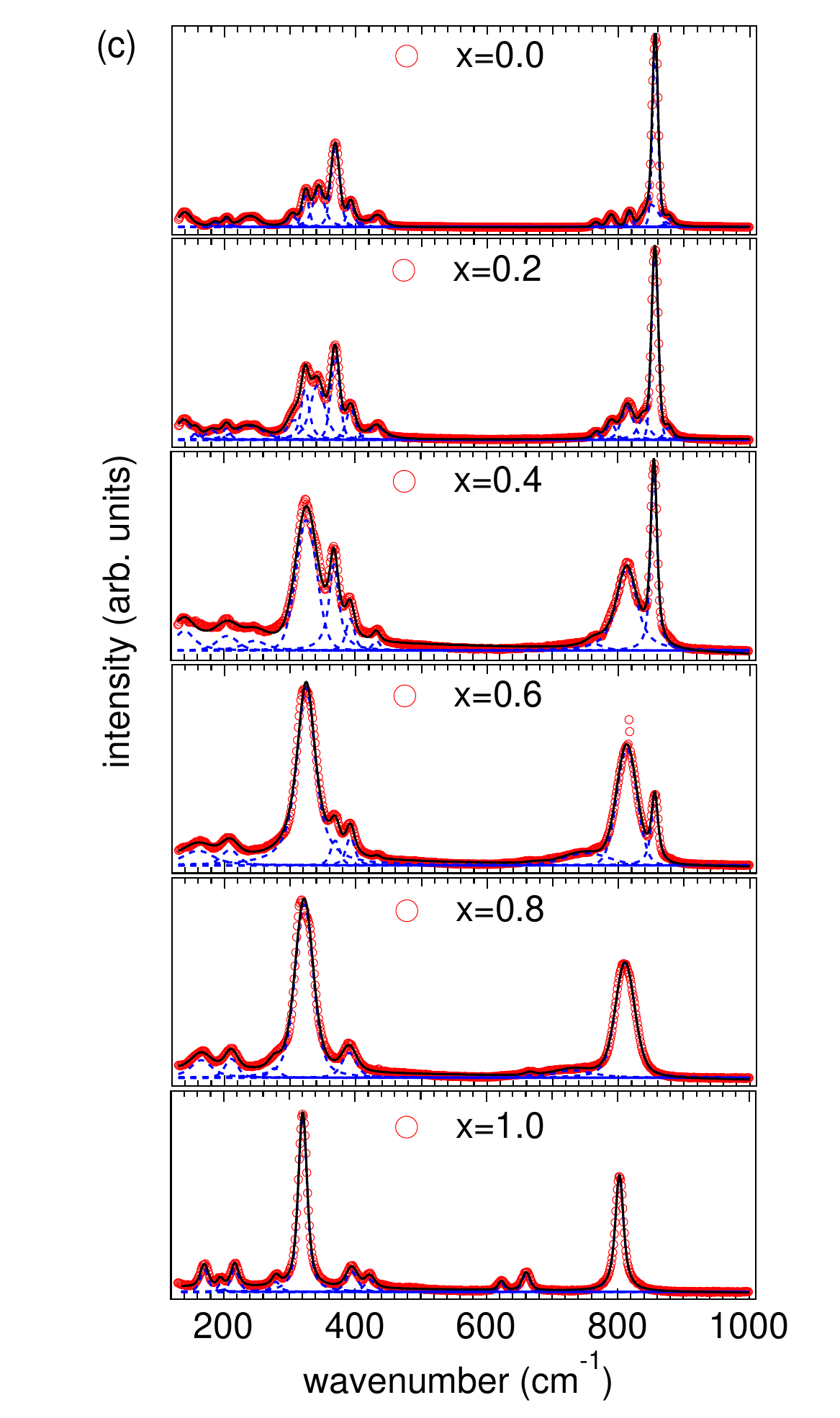} 
\caption{The room temperature Raman spectra of LaV$_{1-x}$Nb$_x$O$_4$ ($x=$ 0 to 1) samples using (a) 532~nm, (b) 633~nm, and (c) 785~nm excitation wavelengths. The dotted blue lines represent the Lorentzian line shape to deconvolute the individual modes.}
\label{fig:Raman}
\end{figure*}

\begin{table*}
\caption{The experimentally observed frequencies $\omega_{obs}$ of the individual Raman modes in LaV$_{1-x}$Nb$_{x}$O$_4$ ($x=$ 0--1) samples measured at room temperature.}
    \centering
\begin{tabular}{|p{0.7cm}|p{1.7cm}|  p{1.7cm}|p{1.7cm}|p{1.7cm}|p{1.7cm}|p{1.7cm}|}

\hline
 $x$ & 0 & 0.2 & 0.4 & 0.6 & 0.8 & 1 \\ \hline
Peak & $\omega_{obs}$ & $\omega_{obs}$ & $\omega_{obs}$ & $\omega_{obs}$ & $\omega_{obs}$ & $\omega_{obs}$ \\ \hline
S$_0$ & B$_g$(127.24) &  &  &  &  & B$_g$(121.86) \\ 
S$_1$ & A$_g$(141.99) & A$_g$(140.66) & A$_g$(140.66) & A$_g$(140.66) &  &  \\ 
S$_2$ & B$_g$(154.05) & B$_g$(155.39) & B$_g$(152.71) &  &  &  \\ 
S$_3$ & A$_g$(187.44) & A$_g$(187.44) & A$_g$(186.11) &  &  &  \\ 
S$_4$ & B$_g$(206.07) & B$_g$(206.07) & B$_g$(206.079) & B$_g$(207.40) & B$_g$(211.39) & A$_g$(219.35) \\ 
S$_5$ & A$_g$(235.26) & A$_g$(233.94) & A$_g$(233.94) &  &  &  \\ 
S$_6$ & A$_g$(245.84) & A$_g$(244.52) & A$_g$(244.52) &  &  &  \\ 
S$_7$ & B$_g$(305.11) & B$_g$(305.11) &  &  &  &  \\ 
S$_8$ & A$_g$(326.07) & A$_g$(326.07) & A$_g$(326.07) & A$_g$(324.76) & A$_g$(328.69) &  \\ 
S$_9$ & A$_g$(344.36) & A$_g$(344.36) & A$_g$(344.36) &  &  &  \\ 
S$_{10}$ & A$_g$(370.41) & A$_g$(370.41) & A$_g$(369.11) & A$_g$(369.11) &  &  \\ 
S$_{11}$ & B$_g$(393.78) & B$_g$(393.78) & B$_g$(391.19) & A$_g$(391.19) & A$_g$(388.60) & B$_g$(396.38) \\ 
S$_{12}$ & A$_g$(420.96) & A$_g$(418.37) & A$_g$(418.37) &  &  &  \\ 
S$_{13}$ & B$_g$(436.44) & B$_g$(435.15) & B$_g$(433.86) & B$_g$(436.44) &  &  \\ 
S$_{14}$ & A$_g$(765.31) & A$_g$(765.31) & A$_g$(766.54) & A$_g$(766.54) &  &  \\ 
S$_{15}$ & B$_g$(788.68) & B$_g$(788.68) & B$_g$(788.68) & B$_g$(788.68) &  &  \\ 
S$_{16}$ & A$_g$(816.86) & A$_g$(815.63) & A$_g$(814.41) & A$_g$(811.96) & A$_g$(807.07) & A$_g$(803.39) \\
S$_{17}$ & A$_g$(840.06) & A$_g$(840.06) & A$_g$(840.06) & A$_g$(840.06) &  &  \\ 
S$_{18}$ & B$_g$(855.88) & B$_g$(855.88) & B$_g$(854.67) & B$_g$(854.67) &  &  \\ 
S$_{19}$ & B$_g$(874.10) & B$_g$(874.10) & B$_g$(874.10) &  &  &  \\ 
S$_{20}$ &  &  &  & A$_g$(170.10) & A$_g$(168.76) & A$_g$(174.10) \\ 
S$_{21}$ &  &  &  &  &  & \\ 
S$_{22}$ &  &  &  &  & A$_g$(108.41) & A$_g$(105.72) \\ 
S$_{23}$ &  &  &  &  &  & B$_g$(164.75) \\ 
S$_{24}$ &  &  &  &  &  & B$_g$(198.09) \\ 
S$_{25}$ &  &  &  &  &  & B$_g$(282.77) \\ 
S$_{26}$ &  &  &  &  & B$_g$(316.91) & A$_g$(322.14) \\ 
S$_{27}$ &  &  &  &  &  & A$_g$(331.30) \\ 
S$_{28}$ &  &  &  &  & A$_g$(345.67) & B$_g$(343.06) \\ 
S$_{29}$ &  &  &  &  &  & B$_g$(405.44) \\ 
S$_{30}$ &  &  &  &  &  & A$_g$(422.25) \\ 
S$_{31}$ &  &  &  &  &  & B$_g$(623.49) \\ 
S$_{32}$ &  &  &  &  &  & A$_g$(648.58) \\ 
S$_{33}$ &  &  &  & B$_g$(669.83) & B$_g$(661.08) & B$_g$(661.08) \\ \hline

\end{tabular}
\label{tab:Ramanmodes}
\end{table*}
\begin{table*}
 \caption{Summary of all the 34 Raman active modes and their assignments with the help of literature (cited in the last coloum of the table) for the LaVO$_4$ and LaNbO$_4$ samples.}

	\centering 	   
\small
\begin{tabular}{|p{0.6cm}|p{1.2cm}|p{5.6cm}|p{1.4cm}|p{6.7cm}|l|}
\hline
       & \multicolumn{2}{c}{LaVO$_4$}   \vline                                                                                             & \multicolumn{2}{c}{LaNbO$_4$}    \vline           &                                                           \\
       \hline
Peak   & \multicolumn{1}{c}{$\omega_{th}$}\vline &  \multicolumn{1}{c}{Assignments}                                                                             \vline & \multicolumn{1}{c}{$\omega_{th}$} \vline &  \multicolumn{1}{c}{Assignments}                                                     \vline          &   \multicolumn{1}{c}{Refs.}  \vline    \\\hline 
S$_0$  & B$_g$(127)                 &  Translation mode of La atoms in monoclinic phase                                           & B$_g$(125.1)               & Coupled translation-rotational mode of La atoms in monoclinic phase and NbO$_4^{3-}$ around an axis perpendicular to b-axis & \cite{IshiiPSSA89,ErrandoneaPMS08,ErrandoneaJPCC16} \\
S$_1$  & A$_g$(143)                 &  Translation mode of La--O bonds                            &                            &                                                   &  \cite{ErrandoneaPMS18,ErrandoneaJPCC16}\\
S$_2$  & B$_g$(158)                 &                Translation mode of La--O bonds        &                            &                          &                        \cite{ErrandoneaPMS18}   \\
S$_3$  & A$_g$(188)                 & Translation mode of La--O bonds                          &                            &                    &                                 \cite{ErrandoneaPMS18,ErrandoneaJPCC16} \\
S$_4$  & B$_g$(204)                 &Translation mode of La--O bonds                               & A$_g$(222.3)               & Translational mode of La--O bonds along b-axis                                    &   \cite{IshiiPSSA89,PellicerJSSC17} \\
S$_5$  & A$_g$(230)                 & Translation mode of La--O bonds                        &                            &                           &                         \cite{ErrandoneaPMS18} \\
S$_6$  & A$_g$(252)                 &Translation mode of La--O bonds                        &                            &                           &                         \cite{ErrandoneaPMS18,ErrandoneaJPCC16} \\
S$_7$  & B$_g$(316)                 & Bending vibration of O--V--O bonds                                                              &                            &                                                     & \cite{HerzbergVNNY87,ErrandoneaJPCC16}\\
S$_8$  & A$_g$(336)                 & Bending vibration of   O--V--O bonds                                                              &                            &                                                    &\cite{HerzbergVNNY87,ErrandoneaJPCC16} \\
S$_9$  & A$_g$(355)                 & Bending vibration of   O--V--O bonds                                                              &                            &                                                      & \cite{HerzbergVNNY87,ErrandoneaJPCC16}\\
S$_{10}$ & A$_g$(380)                 & Bending vibration of   O--V--O bonds                                                              &                            &                                            &          \cite{HerzbergVNNY87,ErrandoneaJPCC16}\\
S$_{11}$ & B$_g$(389)                 & Bending vibration of   O--V--O bonds                                                              & B$_g$(398.1)               & Triply degenerate deformation mode (Rocking mode of   NbO$_4^{3-}$) &
\cite{IshiiPSSA89,HerzbergVNNY87,ErrandoneaJPCC16}\\
S$_{12}$ & A$_g$(423)                 & Bending vibration of   O--V--O bonds                                                              &                            &                                         &           \cite{HerzbergVNNY87,ErrandoneaJPCC16}  \\
S$_{13}$ & B$_g$(427)                 & Bending vibration of   O--V--O bonds                                    &                            &                                                  &    \cite{HerzbergVNNY87,ErrandoneaJPCC16}\\
S$_{14}$ & A$_g$(784)                 & Stretching vibration of V--O bonds &                            &                       &                            \cite{IshaqueSSS99}     \\
S$_{15}$ & B$_g$(799)                 & Stretching vibration of V--O bonds 
&                            &                  &                               \cite{IshaqueSSS99}    \\
S$_{16}$ & A$_g$(806)                 & Stretching vibration of V--O bonds & A$_g$(805.2)               &                                 Non degenerate Stretching mode of  Nb--O bonds 
& \cite{IshaqueSSS99,IshiiPSSA89,HardcastleJPC91,LiuJECS17}\\
S$_{17}$ & A$_g$(836)                 & Stretching vibration of V--O bonds 
&                            &                                               &      \cite{IshaqueSSS99}  \\
S$_{18}$ & A$_g$(861)                 & Non degenerate stretching mode of  shortest V--O bonds &   
&                                                   &  \cite{IshaqueSSS99,ErrandoneaJPCC16} \\
S$_{19}$ & B$_g$(892)                 &Stretching vibration of O--V--O bonds 
&                            &                                                   &  \cite{HardcastleJPC91,LiuJECS17,ErrandoneaJPCC16}\\
S$_{20}$ &                            &                                                                             & A$_g$(177.1)               & Translational mode   along b axis                                &   \cite{IshiiPSSA89}    \\
S$_{21}$ &                            &                                                                            & B$_g$(114)        & Rotational mode of   NbO$_4^{3-}$ along an axis perpendicular to b-axis  & \cite{IshiiPSSA89}\\
S$_{22}$ &                            &                                                                           & A$_g$(108.6)               & Rotational mode of   NbO$_4^{3-}$ along b-axis                       & \cite{IshiiPSSA89,PellicerJSSC17} \\
S$_{23}$ &                            &                                                                           & B$_g$(170)                 & Translational mode   parallel to ac-plane                             & \cite{IshiiPSSA89} \\
S$_{24}$ &                            &                                                                             & B$_g$(200.2)               & Translational mode   parallel to ac-plane                              &  \cite{IshiiPSSA89}\\
S$_{25}$ &                            &                                                                             & B$_g$(284.9)               & Translational mode   parallel to ac-plane                  
& \cite{IshiiPSSA89}\\
S$_{26}$ &                            &                                                                             & A$_g$(321.7)               & Doubly degenerate scissors mode of NbO$_4^{3-}$)
& \cite{IshiiPSSA89,PellicerJSSC17} \\
S$_{27}$ &                            &                                                                            & A$_g$(326.9)               & Doubly degenerate scissors mode of NbO$_4^{3-}$) 
& \cite{IshiiPSSA89,PellicerJSSC17}\\
S$_{28}$ &                            &                                                                             & B$_g$(344)                 & Translational mode   parallel to ac-plane                             &   \cite{IshiiPSSA89}\\
S$_{29}$ &                            &                                                                             & B$_g$(404.9)               & Triply degenerate deformation mode (Rocking mode of   NbO$_4^{3-}$)                                &   \cite{IshiiPSSA89} \\
S$_{30}$ &                            &                                                                             & A$_g$(425.8)               & Triply degenerate deformation mode (Twist mode of NbO$_4^{3-}$)                     &       \cite{IshiiPSSA89}\\
S$_{31}$ &                            &                                                                            & B$_g$(625.5)               & One of triply degenerate Stretching mode of  Nb--O bonds
& \cite{IshiiPSSA89,HardcastleJPC91,PellicerJSSC17}\\
S$_{32}$ &                            &                                                                             & A$_g$(649)                 & One of triply degenerate Stretching mode of  Nb--O bonds          
&\cite{IshiiPSSA89,HardcastleJPC91,PellicerJSSC17} \\
S$_{33}$ &                            &                                                                             & B$_g$(664.9)               & One of triply degenerate Stretching mode of  Nb-O bonds 
& \cite{IshiiPSSA89,HardcastleJPC91,PellicerJSSC17}  \\\hline                            
\end{tabular}
\label{tab:Ramanassignments}
\end{table*}

The Raman spectra of LaV$_{1-x}$Nb$_x$O$_4$ measured at three different excitation wavelengths, 532 nm, 633 nm, and 785 nm are presented in Fig.~\ref{fig:Raman} for all the samples ($x=$ 0--1). Three different excitation wavelengths are used to distinguish the fluorescence effect on the Raman signal and to avoid the background effects from the sample. We use Lorentzian line shape function to deconvolute and fit the observed individual Raman peaks, as marked in Table~\ref{tab:XRD}. We find that all the specific Raman peak positions (Raman shift) are independent of excitation wavelength for a sample, which confirm their inherent characteristic of that particular sample, as shown in Fig.~\ref{fig:Raman}. The intensity of the modes may vary due to several reasons like the polarizability of the molecule, excitation wavelength of the laser source, and the concentration of the active group \cite{ShuklaPRB22}. Though there are minor changes in the intensity variation of Raman modes measured with different excitation wavelengths, we can see that Raman active peaks are changing systematically for all the measured samples in Fig.~\ref{fig:Raman}. 

In the measured spectra we see 20~peaks corresponding to LaVO$_4$ and 17 peaks for LaNbO$_4$. According to the Group theory calculations LaVO$_4$ contains 72 vibrational modes and out of them, 36 modes are Raman active modes (18A$_g$ + 18B$_g$) \cite{SantosJAP07, PanchalPRB11,ChengOM15} (here, A and B denote symmetric and antisymmetric vibrations about the principal axis of symmetry and subscripts $g$ indicates that the vibrations are symmetric relative to a symmetry center, respectively). All the 20 Raman peaks for the $x=$ 0 sample are represented from S$_0$ to S$_{19}$, as shown in the Table \ref{tab:Ramanmodes}. The theoretical approach predicted that the 8A$_g$+10B$_g$  modes are for $m-f$ structure, and 13 Raman-active modes in $t-s$ structure (as observed in the $x=$ 0.6 sample), which are summarized in Table~\ref{tab:Ramanmodes}. The reason for the absence of some of the peaks could be due to the overlap of several A$_g$ and B$_g$ modes and their low Raman scattering cross-section. All the assignments related to each Raman peak in LaV$_{1-x}$Nb$_x$O$_4$ are summarised in Table \ref{tab:Ramanassignments}. We can see in Table \ref{tab:Ramanmodes} that the S$_0$ mode (127.24 cm$^{-1}$) is present only in LaVO$_4$ and LaNbO$_4$ samples and absent for rest of the intermediate samples. The reason for origin of S$_0$ mode is translational motion of La atoms in the monoclinic phase. All the concentrations from $x=$ 0.2 to 0.8 in LaV$_{1-x}$Nb$_x$O$_4$ results in the formation of $t-s$ type structure or $m-m$ and $t-s$ in equilibrium type structure. So, the formation of mixed phase may result in the disappearance of S$_0$ mode. The S$_{18}$ is the most intense mode for the LaVO$_4$ sample, which decreases with Nb substitution. Whereas, we find that the intensity of S$_{16}$ mode increases with Nb substitution and the same becomes the most intense mode for the LaNbO$_4$ sample, as can be seen in Fig.~\ref{fig:Raman}(a). For the $x=$ 0.8 sample, the S$_{18}$ mode completely disappears, which indicates that crystal phase transformation from a mixed phase of $m-m$ and $t-s$ in equilibrium to turning into an approximately pure (96\%) $t-s$ phase \cite{OkramMNL11}. This behaviour of S$_{0}$, S$_{16}$ and S$_{18}$ corroborate with the structural phase transformation with Nb$^{+5}$ substitution, as has been observed in XRD analysis. Furthermore, the presence of S$_8$, S$_9$, S$_{10}$, S$_{13}$, S$_{14}$, S$_{15}$, S$_{17}$ and S$_{18}$ modes in the $x=$ 0 sample confirms the existence of VO$_4^{3-}$ ions since none of these modes are visible in LaNbO$_4$ \cite{ClavierJECS11, SelvanJCS09, HuangN12, WangJL17}. 

All the Raman peaks arise due to different vibrational modes, i.e., bonds between different constituent elements, i.e., the La$^{3+}$, V$^{5+}$, Nb$^{5+}$ and O$^{2-}$. The comparison of experimentally observed peak positions of distinct Raman modes, fitted using Lorentzian function, with the reported data \cite{ErrandoneaJPCC16,JiaEJIC10,IshiiPSSA89,SunJAP10, ChengOM15},  shows a high degree of similarity, as presented in Table~\ref{tab:Ramanmodes}. In the $m-m$ structured LaVO$_4$ crystal, nine O$^{2-}$ atoms are linked to La$^{3+}$ whereas four O$^{2-}$ atoms and V$^{5+}$ are joined in a tetrahedral shape. There are four different O$^{2-}$ locations and it is bound in a 3-coordinate geometry to two equivalent La$^{3+}$ and one equivalent V$^{5+}$ atoms at the first site. Also, it is bound to two comparable La$^{3+}$ and one equivalent V$^{5+}$ atoms in a deformed single-bond geometry in the second site. Three comparable La$^{3+}$ and one equivalent V$^{5+}$ atoms are linked to O$^{2-}$ in a 3-coordinate geometry at the third O$^{2-}$ site and it is bound in a deformed single-bond geometry to three equivalent La$^{3+}$ and one equivalent V$^{5+}$ atom in the fourth O$^{2-}$ site \cite{JainAPLM13}. In the $m-f$ structured LaNbO$_4$ crystal, the La$^{3+}$ is joined to eight O$^{2-}$ atoms in an 8-coordinate geometry and six O$^{2-}$ atoms are bound to Nb$^{5+}$ to create the deformed, edge-sharing NbO$_6$ tetrahedra. There are two different sites for O$^{2-}$ and it is linked in a 4-coordinate geometry to two equivalent La$^{3+}$ and two equivalent Nb$^{5+}$ atoms at the first O$^{2-}$ site. Also, it is bound in a 3-coordinate geometry to two equivalent La$^{3+}$ and one Nb$^{5+}$ atoms at the second O$^{2-}$ site. In the analysis of vibrational modes, it has been assumed that the LaNbO$_4$ crystal is made up of La$^{3+}$ cations and NbO$_4^{3-}$ molecular anions \cite{IshiiPSSA89, JainAPLM13}. It is revealed experimentally that on the addition of Nb$^{5+}$, it replaces V$^{5+}$ from its site and distorts LaVO$_4$’s unit cell \cite{SunCI15}. The mode of vibrations for LaVO$_4$ is categorised as follows: (I) the zone of high wavenumber (765--874 cm$^{-1}$) resulting from O-V-O bond's stretching vibration (II) the intermediate (305--436 cm$^{-1}$) region resulting from O-V-O bond's bending vibration, and (III) the zone of low wavenumber ($<$ 285 cm$^{-1}$) resulting from La atom's translation modes as the La atoms have high mass \cite{SunJAP10,HimanshuPRB21}, and the results are presented in Table~\ref{tab:Ramanassignments}. Similarly, the vibrational modes of LaNbO$_4$ are also categorized as follows: (I) high wavenumber zone (623--803 cm$^{-1}$) for stretching modes of Nb-O bonds, (II) intermediate zone (322--422 cm$^{-1}$) for deformation/scissor modes of NbO$_4^{3-}$, and (III) low wavenumber zone (121--282 cm$^{-1}$) for rotational modes of NbO$_4^{3-}$ and translational lattice modes that include the relative translations of anions and cations \cite{IshiiPSSA89}.

\begin{figure*}
\includegraphics[width=7.1in, height=4.35 in]{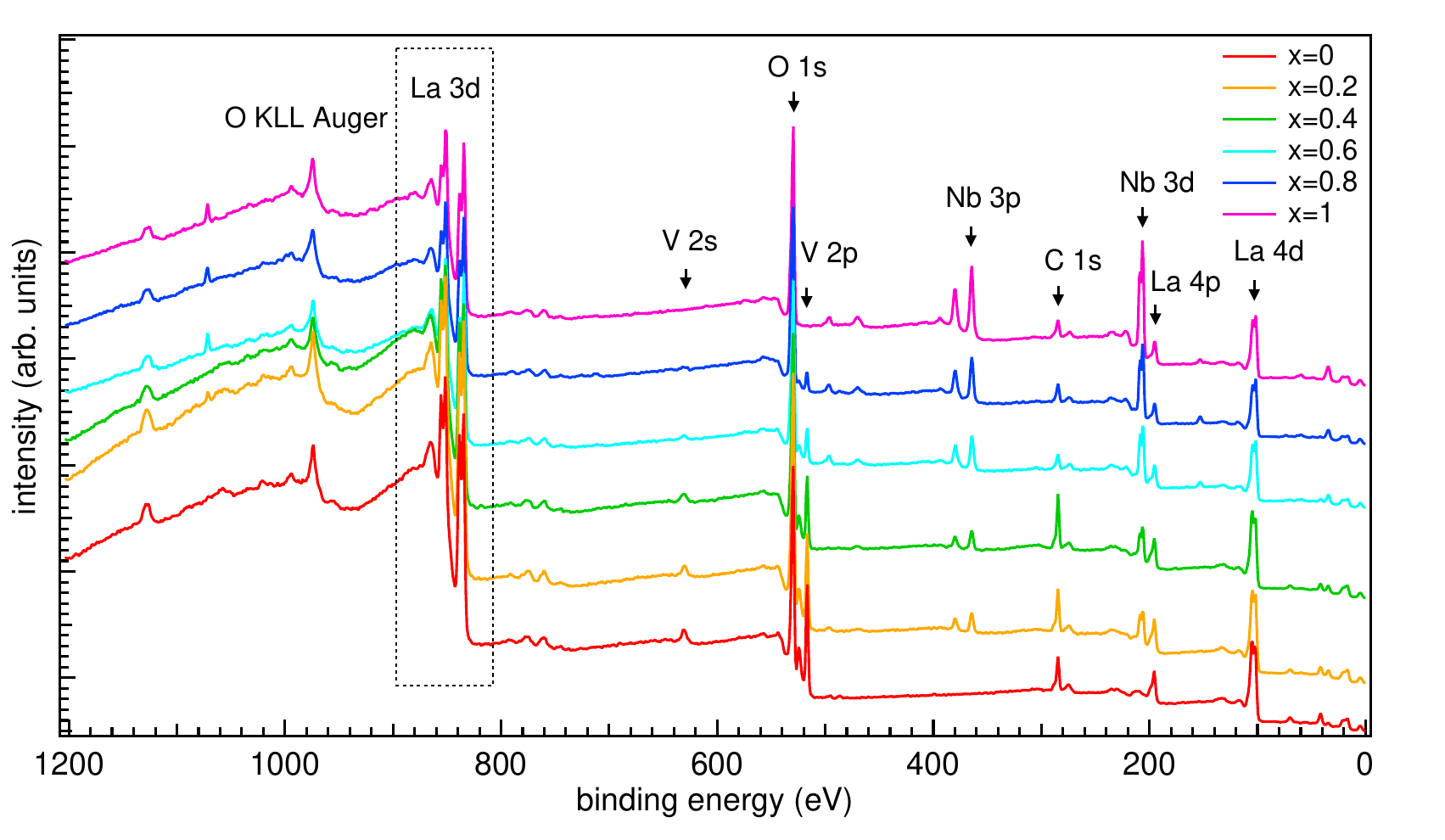}
\caption{The room temperature XPS survey spectra of LaV$_{1-x}$Nb$_x$O$_4$ ($x=$ 0 to 1) samples.}
\label{fig:XPSwide}
\end{figure*}

\begin{figure}
\includegraphics[width=3.6in]{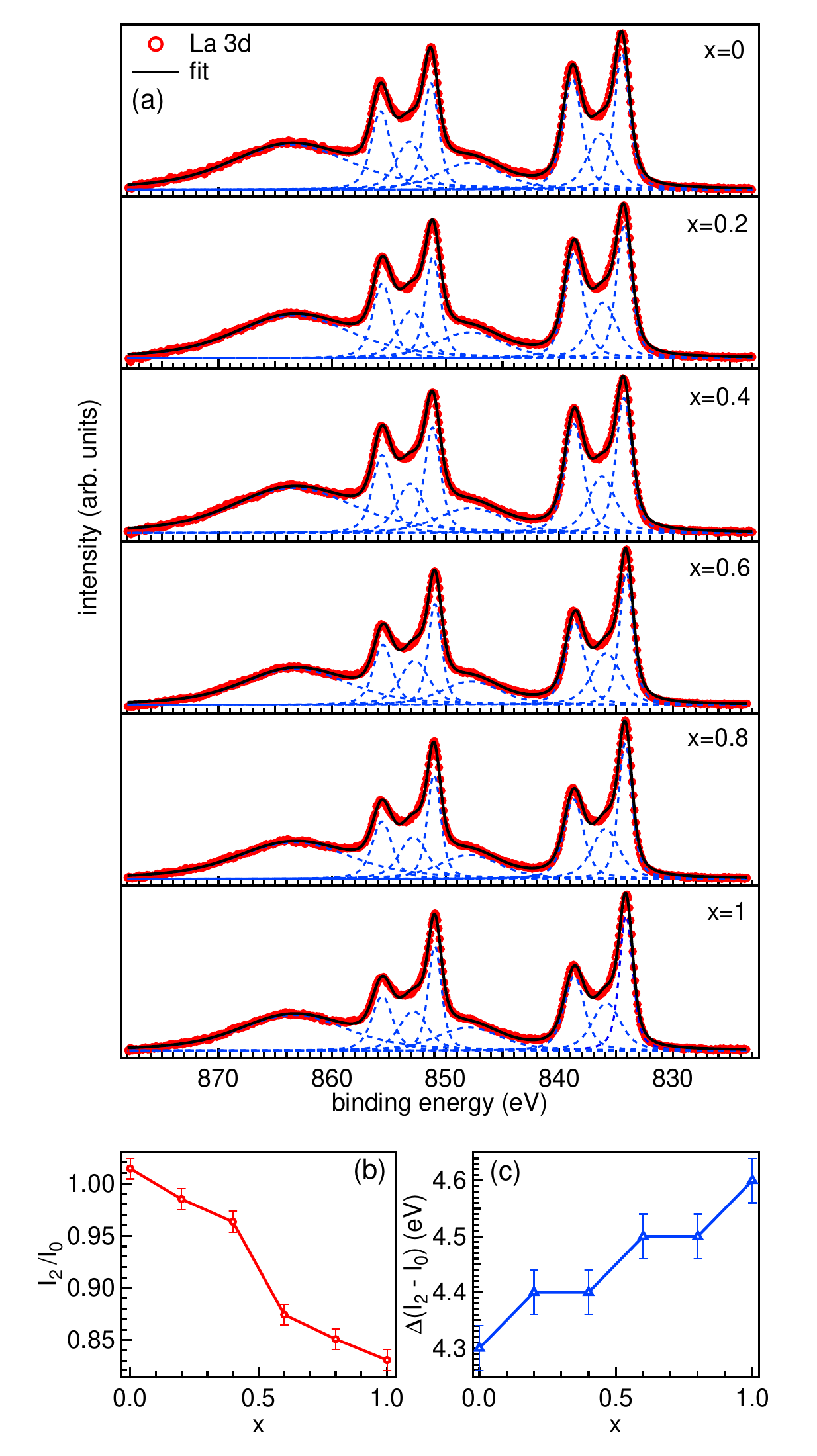}
\caption{(a) The La 3$d$ core-level spectra of LaV$_{1-x}$Nb$_x$O$_4$, $x=$ 0--1 samples. (b) The intensity ratio I$_2$/I$_0$, and (c) energy separation I$_2$ - I$_0$ as a function of doping level $x$. The fitted spin-orbit split components are also shown for each sample.}
\label{fig:XPSLa}
\end{figure}

\begin{figure}[h]
\includegraphics[width=3.4in]{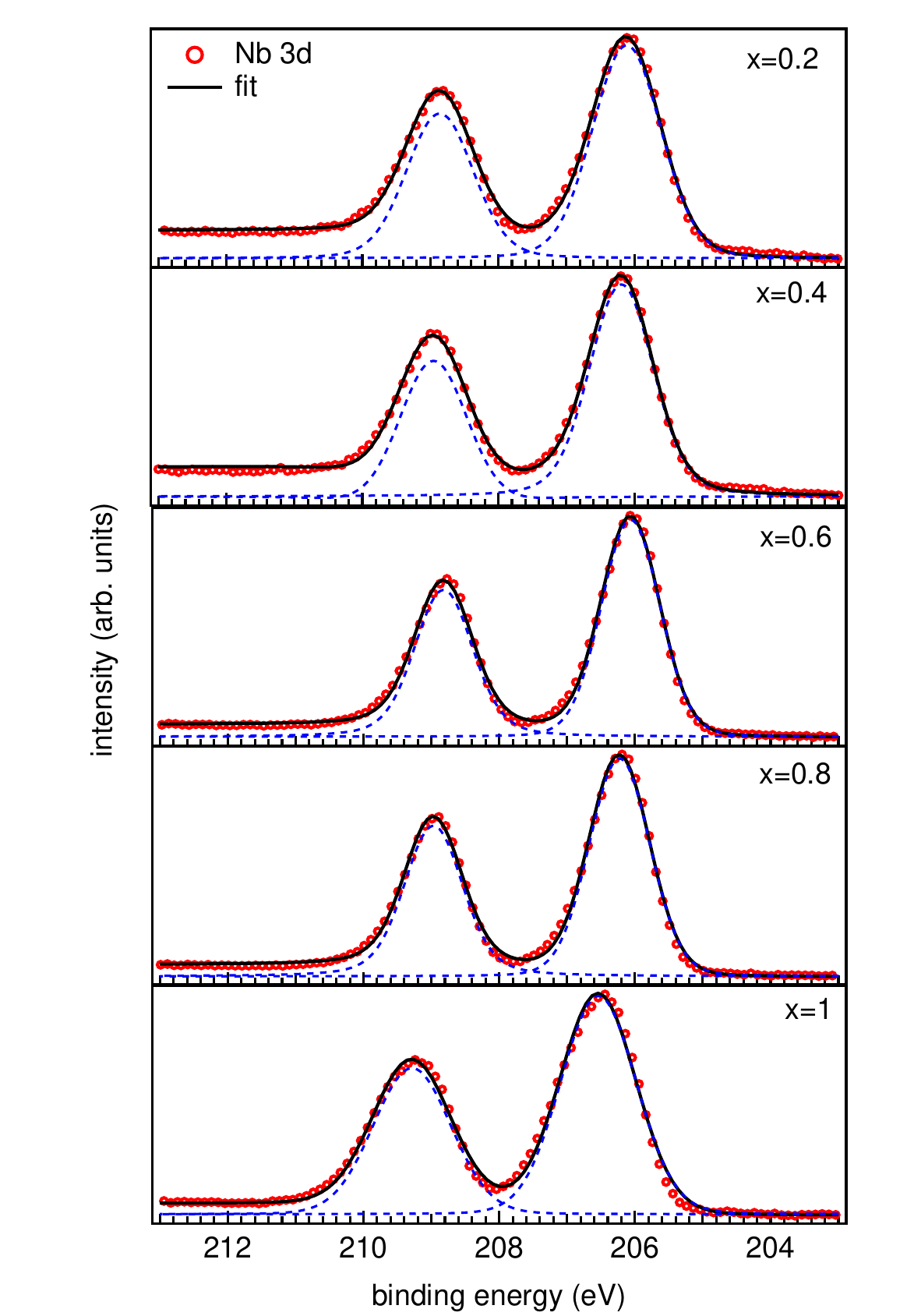}
\caption{Tha Nb $3d$ core level spectra of LaV$_{1-x}$Nb$_x$O$_4$, $x=$ 0--1 samples. The fitted spin-orbit split components are also shown for each sample.}
\label{fig:XPSNb}
\end{figure}
 
The LaNbO$_4$ contains a total of three different modes; rotational modes of NbO$_4^{3-}$, vibrational modes of NbO$_4^{3-}$ and translation modes of La--O and O--La--O bonds. The S$_0$, and S$_{22}$ peaks are visible corresponding to the combined translation-rotational (B$_g$) and rotational mode (A$_g$), respectively, while the third rotational B$_g$ mode (S$_{21}$) is absent in the observed experimental Raman spectra. The vibrational modes can be categorized into (I) doubly degenerated scissor modes, (II) triply degenerated deformation mode, which further splits into a pair of degenerated rocking mode and one twist mode, (III) stretching mode, a non-degenerate and a triply degenerated with increasing order of wave number \cite{IshiiPSSA89}. The remaining modes are all translational modes. From the Table~\ref{tab:Ramanassignments}, we can easily identify that LaNbO$_4$ Raman modes are matching well with the reported one. Two NbO$_4^{3-}$ scissor modes with almost degenerated wave numbers are projected to be seen in the A$_g$ spectrum. Out of all, the most obvious choices are S$_{26}$ and S$_{27}$ because the remaining A$_g$ band's wavenumbers are too low to allocate them. In the LaNbO$_4$, as already discussed, the deformation modes are believed to be divided into two almost degenerated rocking modes (S$_{11}$ and S$_{29}$) with B$_g$ symmetry and a twist mode (S$_{30}$) with A$_g$ symmetry. These modes are also present in the region of intermediate wave numbers. The stretching modes are high energy vibrations and here they are recognised as S$_{16}$, S$_{31}$, S$_{32}$ and S$_{33}$ peaks. As the non-degenerate symmetric mode is expected to provide the strongest band, band S$_{16}$ is allocated to it. The remaining S$_{31}$, S$_{32}$, and S$_{33}$ peaks are assigned to other three degenerate stretching modes. The invariance of the S$_4$, S$_{11}$ and S$_{16}$  peak positions, all through, from the $x=$ 0 to 1 samples indicates no effect on translational mode along the $b-$axis, the B$_g$ frequency of VO$_4^{3-}$ and NbO$_4^{3-}$ for rocking and stretching modes. The S$_8$ peak disappears only in LaNbO$_4$ spectrum because of absence of O-V-O bending vibrations \cite{IshiiPSSA89, HimanshuPRB21}. Interestingly, the S$_2$, S$_3$, S$_5$, S$_6$, S$_9$, S$_{12}$ and S$_{19}$ peaks vanished just before Nb concentration exceeds V (at $x=$ 0.4) and also S$_1$, S$_{10}$, S$_{13}$, S$_{14}$, S$_{15}$, S$_{16}$, S$_{17}$ and S$_{18}$ peaks vanished just after Nb concentration became more than V. It is quite possible that the low concentration of Nb in the sample results in the weakening and then disappearance of some of the spectral peaks. Due to the same reason some new peaks (S$_{20}$ and S$_{33}$) appeared in $x=$ 0.6--1 samples. Furthermore, the S$_{20}$ peak arises due to translational mode along the $b-$axis, and S$_{33}$ peak appeared due to one of three triply degenerate stretching modes of NbO$_4^{3-}$ in the sample.

The most intense peak in LaNbO$_4$ (S$_{16}$) and LaVO$_4$ (S$_{18}$) at higher wavenumber is due to the stretching of Nb-O$_t$  and V-O$_t$ bonds, where O$_t$ represents the oxygen atoms in the terminal position \cite{PenaJSSC20}. The terminal position of oxygen is that where it connects the LaO$_8$ dodecahedra and NbO$_6$ octahedra in case of LaNbO$_4$ and LaO$_9$ muffin \cite{RuizCEJ08} and VO$_4$ tetrahedra in case of LaVO$_4$ \cite{PenaJSSC20}. Since the VO$_4$ tetrahedra appears to be intrinsic in the peak broadening, it has been found that this broadening in the Raman peaks spreads along samples with intermediate Nb and V compositions. However, in certain samples, Nb$^{5+}$ and V$^{5+}$ cation-related variables may also play an important role in the increasing the broadening of the peak. The broad peaks are made up of multiple modes which are normally difficult to distinguish from one another \cite{PenaJSSC20}. The strongest peak of LaVO$_4$ (S$_{18}$) is in the high wavenumber region which lies approximatly 52.5 cm$^{-1}$ on higher side of the spectrum with respect to the strongest peak of LaNbO$_4$ (S$_{16}$). This difference in wavenumber ($\Delta$) is related to average bond length ($d$) of the atoms by $\Delta \propto 1/d^{3/2}$, as stated by the Badger rule \cite{BadgerJCP34}. The bond lengths V--O$_t$ in LaVO$_4$ and Nb--O$_t$ in LaNbO$_4$ are $\sim$1.72~\AA~ \cite{SunJAP10} and $\sim$1.90~\AA~ \cite{PenaJSSC20}, respectively. The changes observed in the Raman sepctra of the samples is quite consistent with the Badger's rule. 

\begin{figure}[h]
\includegraphics[width=3.55in]{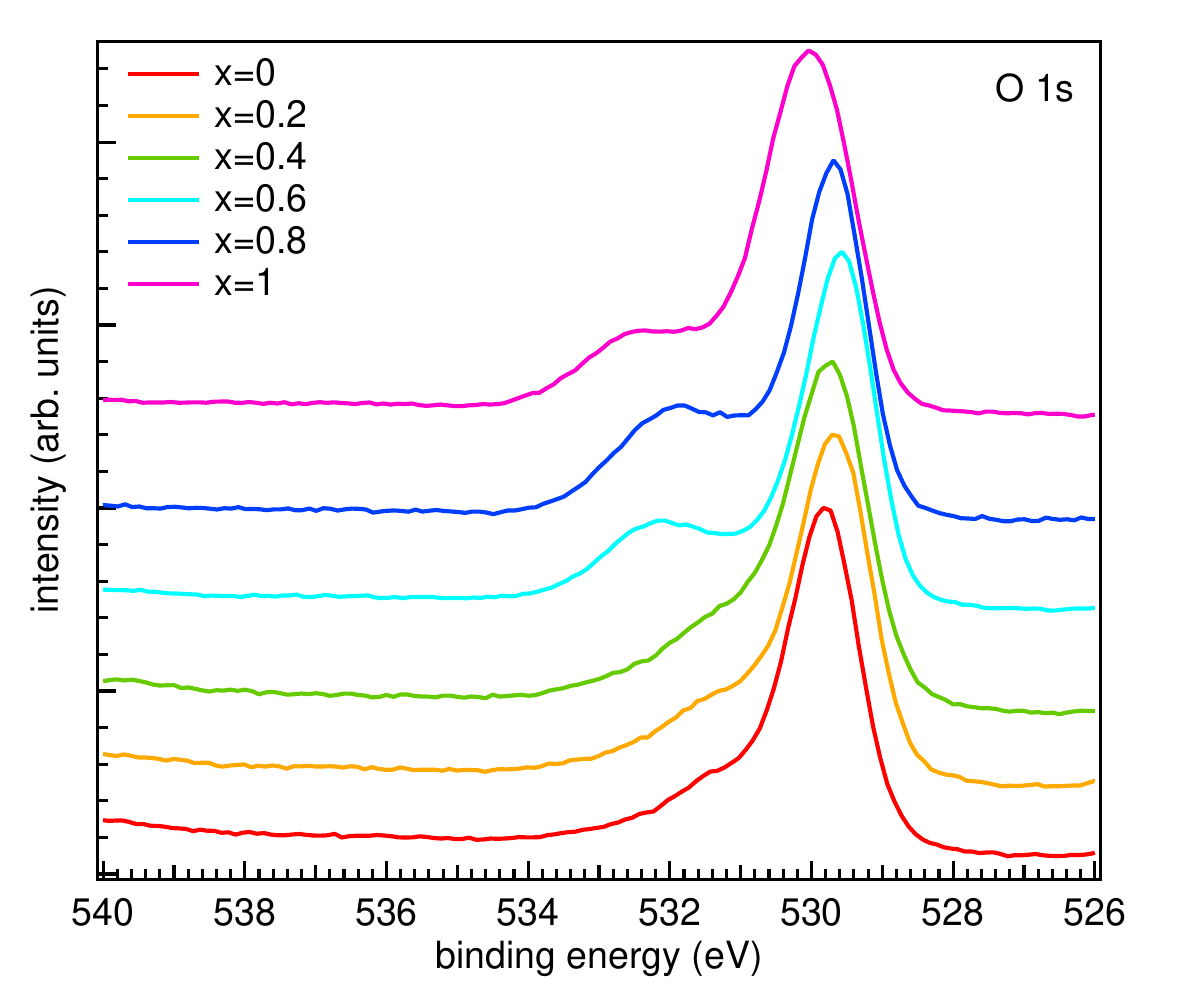}
\caption{Tha O $1s$ core level spectra of  LaV$_{1-x}$Nb$_x$O$_4$, $x=$ 0--1 samples. Each spectrum is shifted vertically for the clarity.}
\label{fig:XPSO}
\end{figure}

\begin{figure}
\includegraphics[width=3.6in, height=5.5 in]{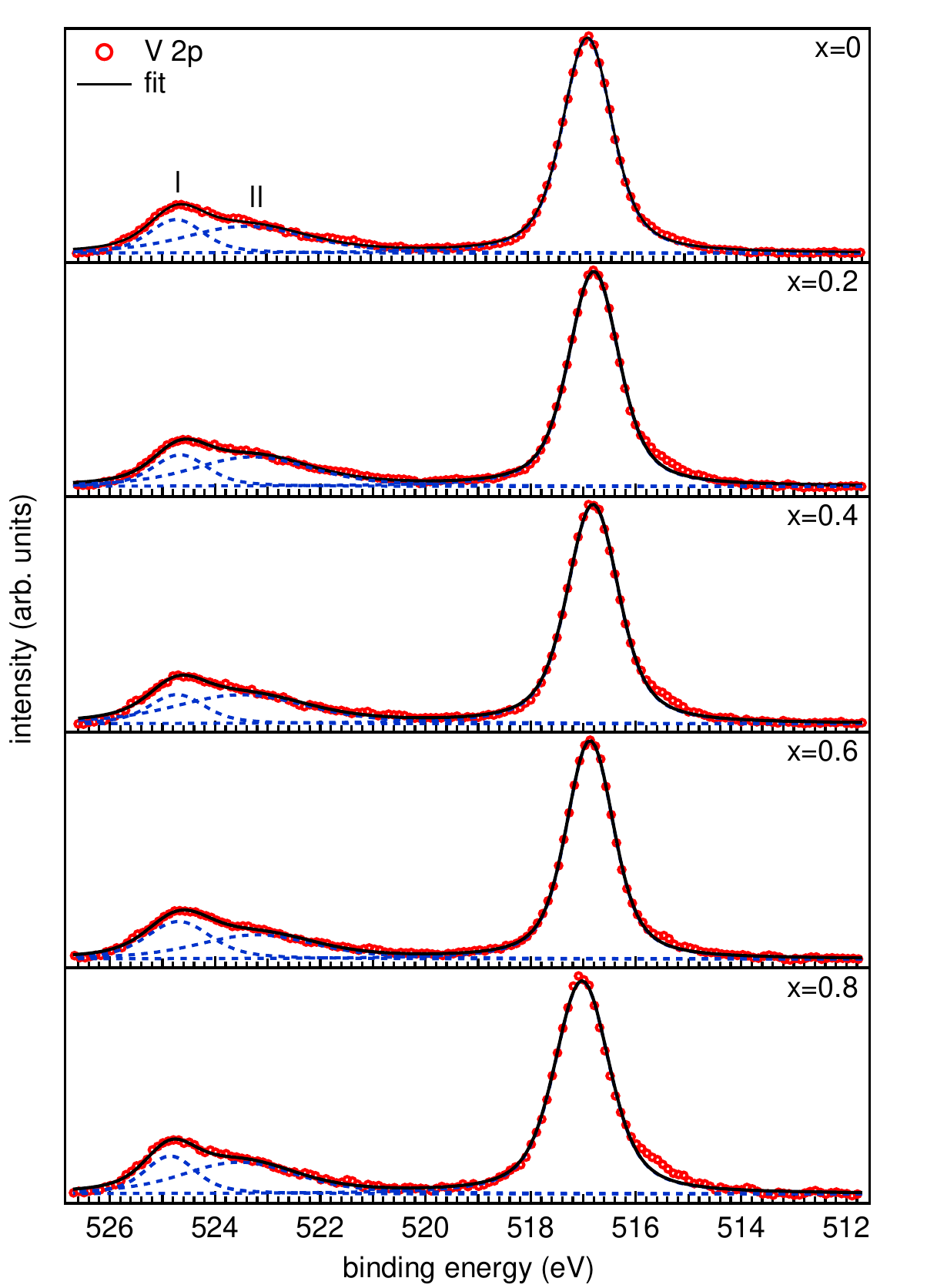}
\caption{Tha V 2$p$ core level spectra of LaV$_{1-x}$Nb$_x$O$_4$, $x=$ 0--1 samples. The fitted spin-orbit split components are also shown for each sample.}
\label{fig:XPSV}
\end{figure}

Finally we use x-ray photoemission spectroscopy (XPS) to investigate the electronic structure  by measuring the survey scan and particular elemental core-level spectra of all the prepared samples. The identified peaks in the survey spectra according to their binding energies are labeled and are in agreement with reported values, as shown in Fig.~\ref{fig:XPSwide}. The characteristic La 3$d$ peaks cluster (830--870 eV), 4$d$ (4$d_{5/2}$ at 101~eV and 4$d_{3/2}$ at 104~eV), and 4$p$ (centered around 195 eV) \cite{Mullica_PRB_85}. The presence of these peaks of La are clearly visible for every synthesised sample, and they are all remarkably comparable. A consistent rise in Nb 3$d$ (discussed later) and Nb 3$p$ (3$p_{3/2}$ at 364 eV and 3$p_{1/2}$ at 379 eV) is observed along with an increase in Nb doping and this feature of Nb is absent in the $x=$ 0 sample \cite{Steiner_ZPB_79}. For the V 2$p$ (2$p_{3/2}$ at 517 eV and 2$p_{1/2}$ at 525 eV) and V 2$s$ (630 eV) core-level peaks, the reverse behavior is anticipated and it is quite evident as clearly visible in Fig.~\ref{fig:XPSwide} \cite{Lebugle_PS_81}. The Voigt function has been used to fit the core level spectra of the constituent elements. The fitted La 3$d$ core-levels are shown in Fig.~\ref{fig:XPSLa}(a). The spin-orbit splitting peaks present in all the samples, have been de-convoluted at binding energies 834.3$\pm$0.2 eV, 836.0$\pm$0.3 eV, 838.7$\pm$0.1 eV, 847.9$\pm$0.1 eV, 851.1$\pm$0.2 eV, 853.0$\pm$0.3 eV, 855.6$\pm$0.1 eV, and 863.4$\pm$0.2 eV (average B.E. of all the samples $\pm$ $\Delta$B.E., calculated for the $x=$ 0--1 samples). The broad diffusive satellite peaks at 847.7 eV and 863 eV in the locality of La 3$d$ core-level are coming from plasmons. Because of the two final states I and II, and the subsequent spin-orbit splitting between each state, making the structure complex. The primary strong peaks (3$d_{5/2}$ at 834.3 eV and 3$d_{3/2}$ at 851.1 eV, respectively) are associated with the final state I (La$^{4+}$ 3d$^9$4f$^0$, L), which involves electron transfer to the continuum from the 3$d$ core-level. The peaks at higher binding energies are features of final state II (La$^{3+}$ 3d$^9$4f$^1$, L, -e) and this feature is experimentally unresolved which indicates multiplet structure, as has been suggested by Mullica {\it et al.} \cite{Mullica_PRB_85}. This corresponds to the electron transfer from ligand (L, O$_{2p}$ in our case) valance band to the empty 4$f$ orbitals of La \cite{Mullica_PRB_85, Shukla_JPCC_19}. This multiplet structure of state II is composed of two bonding and anti-bonding states. The prominent signals at higher binding energies (3$d_{5/2}$ at 838.7 eV and 3$d_{3/2}$ at 855.6 eV) are due to bonding of state II and the weak signals at lower binding energies (3$d_{5/2}$ at 836.0 eV and 3$d_{3/2}$ at 853 eV) are because of anti-bonding. The average energy difference (over La core-level spectra of all the samples) between these three pairs of peaks is nearly the same (~16.9 eV) for the state I, state II bonding, and state II anti-bonding, respectively. This verifies the unaltered spin-orbit energy splitting of the states of La on Nb substitution \cite{Shukla_PRB_2023}. Interestingly, we find a significant and systematic change in the intensity variation of the peak at 838.7 eV (I$_2$) relative to the primary peak at 834.3 eV (I$_0$) with Nb doping. The metal-ligand orbital overlaps are reported to be accountable for such doping-induced intensity variations \cite{Kamath_IJC_1984, Vasquez_PRB_1996} where strong ligands are found to populate the (La$^{3+}$ 3d$^9$4f$^1$, L, -e) state, intensifying I$_2$ \cite{Signorelli_PRB_1973}. The intensity ratio I$_2$/I$_0$ is shown in Fig.~\ref{fig:XPSLa}(b), which shows a consistent decrease as a function of doping $x$. This signifies that with Nb substitution, the extent of overlapping between La(4f)-O(2p) orbitals decreases monotonically. This conclusion can also be drawn from the trend in the energy separation between I$_2$ and I$_0$ as a function of $x$, as shown in Fig.~\ref{fig:XPSLa}(c). However, the separation is minute in the subsequent samples, but for the $x=$ 0 and $x=$ 1 the energy difference (I$_2$ - I$_0$) is found to be of the order of 0.3 eV. The value of I$_2$ - I$_0$ was found to be varying for a variety of La-containing compounds majorly because of the crystal structure, like 3.8 eV for La$_{0.5}$Sr$_{0.5}$Co$_{1-x}$Nb$_x$O$_3$  and 5.3 eV La$_{1.85}$Ba$_{0.15}$CuO$_4$ \cite{Shukla_PRB_2023, Vasquez_PRB_1996}. Notably, this energy separation could be related to the ease of electron transfer between the ligand and the more ionic state of La, therefore having an opposite trend with the tendency of ligand’s overlapping with the La $4f$ orbitals \cite{Vasquez_PRB_1996}.

The Nb 3$d$ core level spectra are shown in Fig.~\ref{fig:XPSNb} where the spin-orbit doublet of Nb 3$d$ core levels are fitted with a single peak for each component and the calculated peak positions for the Nb-doped samples are found to be 3$d_{5/2}$ at 206.2$\pm$0.2 eV and 3d$_{3/2}$ at 209.0$\pm$0.2 eV \cite{Shukla_PRB_18, Isawa_PRB_94}. This confirms the presence of prevailed 5$+$ oxidation state of Nb atom \cite{Steiner_ZPB_79} in all the samples. However, for the $x=$ 1 sample the Nb 3$d_{5/2}$ is at a higher binding energy as compared to the other Nb-containing samples, which could be due to the charging effects and the change in chemical environments. Therefore, Atuchin \textit{et al.} characterized the Nb state by using energy difference $\Delta$ (Nb 3$d_{5/2}$ -- O 1$s$) instead of solely relying on Nb 3d$_{5/2}$ binding energy position \cite{Atuchin_JESRP_05}. The evaluated $\Delta$ (Nb 3$d_{5/2}$ -- O 1s) values are found to be around 323.5 eV. The calculated energy difference with respect to O 1$s$ is independent of the carbon correction. The obtained binding energy difference $\approx$323.5 eV is reported to be a fairly highest value for the 5+ oxidation state of Nb. We can also see that the error in the value of binding energy position $\Delta$ is only 0.1 eV in this case, while for the Nb 3d$_{5/2}$, and O 1$s$, it is 0.3, and 0.2 eV, respectively. In Fig.~\ref{fig:XPSO} we can also see that the O 1$s$ peak is shifting to higher binding energy for the $x=$ 1 sample as compared to the $x=$ 0. Similarly, for the Nb 3$d_{5/2}$ core-level, it is shifting to higher binding energy. The $\Delta$(Nb 3d$_{5/2}$ - O 1s) value for the $x=$ 1 sample is quite consistent for all the samples, which strongly supports the electronic characterization using energy difference with respect to O 1$s$ instead of absolute peak positions. 

In Fig.~\ref{fig:XPSV}, we present the V $2p$ core level spectra for all the samples, which shows spin orbit components of 2$p_{3/2}$, and 2$p_{1/2}$ at 516.9 and 524.8 eV, respectively indicating V in ${5+}$ state. Interestingly, an unusual broadening in the V $2p_{1/2}$ component is observed for all the samples, whereas no such additional component is evident in the V $2p_{3/2}$ peak at 516.9 eV. More importantly, the deconvolution of the V $2p_{1/2}$ component reveals that the FWHM of the higher energy feature (denoted by I) (1.2~eV) is nearly the same as that of $2p_{3/2}$ component (1.1~eV). In contrast, the lower energy feature (II) is significantly broader (2.8~eV). Moreover, the area ratio of the combined I and II with $2p_{3/2}$ is close to 1/2, which clearly indicates the intrinsic origin of these two features from the vanadium. In contrast to metallic V $2p$ core-level, the vanadium based compounds have often been reported to exhibit an anomalous V $2p_{1/2}$ width as a consequence of Coster-Kronig (C-K) transitions \cite{Sawastzky_PRB_79}. The C-K type transition is a class of Auger transition in which an electron from a higher sub-shell of the same shell fills the core hole \cite{Antonides_PRB_77}. In the present case, the filling of $2p_{1/2}$ core hole by an electron from $2p_{3/2}$ may give rise to the C-K transitions and that can result in an additional feature in the $2p_{1/2}$ component. Therefore, it is likely that the component I is attributed to the core-hole recombination with the screening electrons, analogous to the 2p$_{3/2}$, whereas an additional L$_2$-L$_3$ (C-K) relaxation process gives rise to the feature II in 2p$_{1/2}$ peak \cite{Ohno_JESRP_04}. No significant change in these components has been observed with the Nb substitution, indicating the robust nature of the underlined system. Further, the approach of O $1s$ energy difference is also implemented in this case as for vanadium oxides, the energy difference $\Delta$(V 2p$_{3/2}$ - O 1s) is an advantageous reference \cite{Mendialdua_JESRP_95}. The average $\Delta$(V 2p$_{3/2}$ - O 1s) magnitude is 12.8$\mp$0.1 eV, which is in good agreement with the literature for V$^{5+}$ oxidation state \cite{Silversmit_JESRP_04}. 

\section*{\noindent ~Conclusions}

In conclusion, the solid state reaction method was used to prepare LaV$_{1-x}$Nb$_x$O$_4$ samples successfully with regular variable Nb$^{5+}$ concentration. The XRD measurements established that the substitution of larger Nb$^{5+}$ ion for V$^{5+}$ affects the lattice constant of LaV$_{1-x}$Nb$_x$O$_4$ and goes through three different phase transformations [monoclinic monazite ($m-m$) type ($x=$ 0), two-phase equilibrium of monoclinic monazite ($m-m$) and tetragonal scheelite ($t-s$) type (0.2$\leq$$x$$\leq$0.8) and monoclinic fergusonite ($m-f$) type ($x=$ 1)]. The SEM micrographs helped in analyzing that the particle size and shape altered due to the change in crystal phases of these samples with increasing Nb$^{5+}$ concentration. The analysis of HR-TEM and SAED data found to be consistent with the XRD refinement data. The Raman spectra of LaV$_{1-x}$Nb$_x$O$_4$ were studied using 532 nm, 633 nm, and 785 nm excitation wavelengths. All the Raman assignments were found to have well-ordered enhancement/diminution with the increase in Nb$^{5+}$ doping. The variation in the intensity as well as appearance/disappearance of the Raman mode with Nb concentration are coinciding with the change in the structural phases, as observed in XRD analysis. This further confirms that the phase transformation in LaV$_{1-x}$Nb$_x$O$_4$ agrees with the maximum intensity peak patterns shown in the Raman spectra of these samples and are consistent with the Badger's rule. The XPS analysis reveal the changes in Nb 3$d$ and V 2$p$ core-level spectral intensities of the samples with the increase in Nb$^{5+}$ concentration. The equal spin-orbit energy splitting of the states was confirmed by the average energy difference (over La core spectra of all samples) for state I, state II bonding, and state II anti-bonding and the observed changes in their relative intensities with Nb substitution are due to the metal ligand orbitals overlap. These findings provide valuable insights into the structural and electronic properties of LaV$_{1-x}$Nb$_x$O$_4$ samples and their potential use in different fields of practical applications. 




\section*{\noindent ~Acknowledgment}
AS and MS thank MHRD and CSIR, respectively for the fellowship. The authors acknowledge IIT Delhi's FIST (DST, Govt. of India) UFO scheme for providing the physics department with the Raman facility. We thank the Physics Department at IIT Delhi for the XRD and the Central Research Facility (CRF) for the FESEM, EDX, and XPS. We also thank Ambuj Mishra for providing HR-TEM facility at IUAC, New Delhi. The preparation of the samples was done in a high-temperature furnace (from Nabertherm GmbH, Germany), funded by BRNS through the DAE Young Scientist Research Award (Project Sanction No. 34/20/12/2015/BRNS). RSD acknowledges SERB--DST for the financial support through a core research grant (project reference no. CRG/2020/003436).

\end{document}